\documentclass[prl,twocolumn,showpacs,showkeys]{revtex4}

\usepackage{amssymb, amsmath, amsfonts, latexsym, verbatim, mathrsfs}

\usepackage{graphicx}

\begin{document}

\title{Quantum and Superquantum Nonlocal Correlations}
\thanks{One of us, R.R., acknowledges the financial support by the R. Parisi Foundation }

\author{GianCarlo Ghirardi}
\email{ghirardi@ictp.it}
\affiliation{Department of Physics, University of Trieste, the Abdus Salam ICTP, Trieste and INFN, Sezione di Trieste, Italy}
\author{Raffaele Romano}
\email{rromano@ts.infn.it}
\affiliation{Department of Physics, University of Trieste, Fondazione Parisi, Rome, Italy}


\begin {abstract}
\noindent We present a simple hidden variable model for the singlet state of a pair of qubits, characterized by two kinds, hierarchically ordered, of hidden variables.
We prove that, averaging over both types of  variables, one reproduces all the quantum mechanical correlations of the singlet state. On the other hand,  averaging only over
the hidden variables of the lower level, one obtains a general formal theoretical scheme  exhibiting  correlations stronger than the quantum ones, but with faster-than-light communication forbidden.
This result is interesting by itself since it shows that a violation of the quantum bound for nonlocal correlations can be implemented in a precise physical manner and not only mathematically, and it suggests that resorting to two levels of nonlocal hidden variables might led to a deeper understanding of the physical principles at the basis of quantum nonlocality.
\end{abstract}




\maketitle


{\it Introduction ---} Nonlocal correlations have always been considered as one of the peculiar traits of quantum
mechanics. Their relevant conceptual implications have been already recognized by the founders of the quantum
theory, as clearly expressed by Schr\"{o}dinger \cite{Erwin}:
\begin{quote}
{\it It is rather discomforting that the theory should allow a system to be steered or piloted into one or the other type of state at the
experimenter's mercy in spite of his having no access to it.}
\end{quote}
Their potential impact for technological implementations is at the basis of the theories of quantum information, computation
and cryptography. Nonetheless, the physical principles underlying quantum nonlocality are rather obscure (see \cite{Pawlowski} for
recent developments).

One of the most relevant features of quantum correlations is that, while they can produce instantaneous actions at a distance \cite{Einstein},
they cannot be used for faster-than-light communication. This no-signalling property allows quantum mechanics to peacefully
coexist with relativistic causality. Nevertheless, such a property is also consistent with generalized models which are more nonlocal than quantum
mechanics \cite{Popescu}. Accordingly, one cannot account for   the specific form of quantum nonlocality in terms of causality requirements. The investigation of the superquantum correlations
generated by these models is important from an information theoretic point of view, and it is expected to lead to a deeper understanding of the physical principles at the basis of quantum mechanics \cite{Barrett,Cerf,Skrzypczyk}. In this perspective, any connection between generalized no-signalling
models and quantum mechanics might represent a relevant conceptual step for the understanding of the origin and of the role  of nonlocality in the quantum
domain.

Recently, Leggett exhibited a no-signalling hidden variable model of a new type \cite{Leggett}, which stimulated investigations on the nature
of quantum nonlocality \cite{Gisin, Groblacher, Colbeck, Parrott}. In this paper we present a simple hidden variable model for the singlet of a two-qubits system, which is strongly related to the model described by Bell in his analysis of nonlocality \cite{Bell}, and constitutes an example of the theories considered by Leggett. Although it is just a toy model, it throws an interesting light on the general problem of nonlocality, since it provides a meaningful connection between the main ideas of \cite{Popescu,Leggett}. In fact, it can produce nonquantum nonlocal correlations, which  reduce to the the standard quantum ones when integration over all hidden variables is performed. This emergence of quantum correlations from
a generalized no-signalling model was partially obtained in a former work \cite{Skrzypczyk}, without reference to hidden variables theories.

In the following we consider a bipartite system shared between two spatially separate observers, which can independently choose inputs
${\bf a}$ and ${\bf b}$, and obtain outputs $a$ and $b$. In general, ${\bf a}$ and ${\bf b}$ are the physical setting which separately
determine the local observables $A$ and $B$, and $a$, $b$ are the respective measurement results.

\noindent We start by briefly reconsidering the arguments of \cite{Popescu,Leggett}.

{\it The analysis of Popescu and Rohlrich ---} To illustrate the arguments of these authors we follow their line of thought. Let $A,A',B$
and $B'$ be physical variables taking values $+1$ and $-1$, with $A$ and $A'$ referring to measurements on one part of the system and $B$
and $B'$ referring to the other part. If we denote as $P_{AB}(a,b)$ the joint probability of obtaining $A=a$ and $B=b$ when both $A$ and
$B$ are measured, the correlation $E(A,B)$ of the outcomes is defined as:
\begin{eqnarray}
E(A,B) &=& P_{AB}(+1,+1) + P_{AB}(-1,-1) + \\ \nonumber
&-& P_{AB}(+1,-1) - P_{AB}(-1,+1).
\end{eqnarray}
As well known Clauser, Horne, Shimony and Holt \cite{CHSH} have shown, completely in general, that an appropriate combination of correlations
involving four arbitrary directions satisfy, for all local theories, the inequality $\vert F \vert\leq{2}$, where
\begin{equation}\label{f1}
    F \equiv E(A,B) + E(A,B') + E(A',B) - E(A',B').
\end{equation}

On the other hand, in the quantum case, when the system is in an entangled pure state, the correlations $E_{\psi}(A,B) \equiv \langle \psi
\vert A \otimes B \vert \psi \rangle$ violate, for appropriate choices of the observables, the above inequality. Actually, Bell \cite{Bell}
 has derived his celebrated inequality, $\vert F_{\psi} \vert \leqslant 2\sqrt{2}$, where $F_{\psi}$ has the same form of (\ref{f1}) with
 $E_{\psi}(X,Y)$ replacing $E(X,Y)$. It has to be noted that the value $2\sqrt{2}$ represents the maximal possible violation of locality which can occur within quantum mechanics.

The authors of \cite{Popescu} have investigated whether the request that the hypothetical general nonlocal theory one is envisaging respects
relativistic causality might be responsible for the precise value of Bell's upper bound. The question is interesting since, at first sight,
one might expect that the above combination of correlations reaches the value $4$, which is attained when the first three terms take the
value $+1$ and the last the value $-1$ (or viceversa). The extremely interesting result of \cite{Popescu} is that a nonlocal theory respecting
relativistic causality and yielding a violation of  Bell's bound is actually possible. We will call any theory exhibiting such a feature a
superquantum nonlocal theory. Therefore, Bell's bound does not follow from a fundamental principle (preservation of causality), but it is rather
a consequence of the structure of quantum mechanics.

The prototype of a device which is more nonlocal than quantum mechanics, still consistent with relativistic causality, is the so-called {\it
Popescu-Rohrlich (PR) box} \cite{Barrett,Cerf}. Its inputs, denoted here by ${\bf a}$ and ${\bf b}$ for consistency, independently chosen by Alice
and Bob respectively, are classical bits which assume the values $\{0, +1\}$. Its outputs, denoted by $a$ and $b$ respectively, are classical bits
taking the same values, determined by the relation ${\bf a b} = a + b$, where the sum is modulo 2. Alice inputs ${\bf a}$ and extracts $a$, and
Bob inputs ${\bf b}$ and extracts $b$. If the outputs are redefined as $a' = 1 - 2a$, $b' = 1 - 2b$, such that the outcomes are $\{-1, +1\}$,
it is possible to check that the PR box provides correlations satisfying $\vert F \vert = 4$. To our knowledge, the only possible
implementation of the PR box so far discussed requires post-measurement selection \cite{Marcovitch}.

{\it Leggett's proposal ---}
Leggett embraces the perspective that quantum mechanics is an incomplete theory, and argues that the hidden variables required to complete it could enable a deeper understanding of quantum correlations. The hidden variables fully specify the values assumed by the physical
quantities in a quantum state $\psi$, denoted by $a_{\psi}$ and $b_{\psi}$; moreover, Leggett assumes that the violation of locality does respect
parameter independence, i.e. that while the  value of $a_{\psi}({\bf a},{\bf b},\lambda)$ might depend on both settings ${\bf a}$ and ${\bf b}$,
it does not depend on the value taken by $B$ \footnote{Actually, this request is necessary because, as well known, deterministic hidden variable
theories can violate Bell's locality condition only by violating Parameter Independence, i.e. by allowing that the outcome of a measurement on
$A$ can (and in general actually it does) depend on the fact that also $B$ suffers a measurement, but not on its outcome.}. As usual, the
ignorance on the specific values of the hidden variables is responsible for the statistical content of quantum mechanics. It is a well known
fact that all single and joint outcomes implied by a maximally entangled state cannot be described in terms of a classical distribution over
the set of hidden variables. Nevertheless, Leggett has investigated wether  it would be possible - at least in principle - to have a local
description of local outcomes when only a subset of the hidden variables is taken into account to evaluate the averages. He has characterized his attempt as taking into account {\it crypto nonlocal theories}.

The ideas of Leggett, originally introduced in the description of the rotationally invariant maximally entangled state of the polarization
degrees of freedom of a pair of photons, are summarized here in terms of a more general theory. The hidden variables are given
by $\lambda = (\mu, \tau)$, where $\mu$ and $\tau$ are two families of hidden variables. It is assumed that
\begin{eqnarray}\label{leggett}
    \langle A \otimes B \rangle_{\psi} &=& \int E_{\psi, \tau} (A,B) \, \rho(\tau) \, d \tau, \nonumber \\
    \langle A \rangle_{\psi} &=& \int f_{\psi} ({\bf a},\tau) \, \rho(\tau) \, d \tau,
\end{eqnarray}
and similarly for $\langle B \rangle_{\psi}$, with $ g_{\psi} ({\bf b},\tau) $ taking the place of  $f_{\psi} ({\bf b},\tau)$. In this equation,
the intermediate averages over $\mu$ are given by
\begin{eqnarray}\label{intave}
  E_{\psi, \tau} (A,B) &=& \int a_{\psi} ({\bf a},{\bf b},\mu,\tau)  b_{\psi} ({\bf a},{\bf b},\mu,\tau) \, \rho(\mu|\tau) \, d\mu, \nonumber \\
  f_{\psi} ({\bf a},\tau) &=& \int a_{\psi} ({\bf a},{\bf b},\mu,\tau) \, \rho(\mu|\tau) \, d\mu, \\
  g_{\psi} ({\bf b},\tau) &=& \int b_{\psi} ({\bf a},{\bf b},\mu,\tau) \, \rho(\mu|\tau) \, d\mu, \nonumber
\end{eqnarray}
The constraints on local measurements, embodied by the second and third of (\ref{intave}), provide the fundamental condition imposed to the model, denoted by Leggett as {\it
crypto-nonlocality}, i.e., that averaging the single particles values for $A$ and $B$ over the "deep level" hidden variables $\mu$, nonlocality
is washed out. This is a no-signalling condition.

Leggett himself has proved that his proposal conflicts with some predictions of quantum mechanics. Moreover, his
work has stimulated a series of interesting  investigations \cite{Gisin, Groblacher, Colbeck, Parrott},
proving in particular that any crypto-nonlocal theory equivalent to quantum mechanics must be characterized by $f({\bf a},\tau) = 0$ and $g({\bf b},\tau) =0 $.

These conclusions might be (and actually have been)  taken as a proof that it is  useless to consider crypto-nonlocal theories (i.e. theories with two levels of the
hidden variables) since they would, in practice, turn out to be a sort of local theories which are already known to be incompatible with quantum predictions. However, as we are going to prove, this is not the case, because it is possible to work out theories of this type which reproduce the quantum single
particle expectation values when integrated on the deeper level hidden variables, but which require also the integration on the upper level
variables in order to reproduce the quantum correlations, while, at the lower level, they violate the quantum upper bound of $2\sqrt{2}$
on the combination of the correlations. In brief, at their lower level such theories are superquantum nonlocal.

{\it Our proposal ---} We take inspiration by the simple nonlocal hidden variable model introduced by Bell in his celebrated paper. However,
we improve it by making it more symmetric for what concerns its nonlocal features, and we modify it in order to make it of the crypto-nonlocal
type. The model describes a pair of qubits in the singlet state $\psi$ in terms of a hidden variable, which is a unit vector $\lambda$ in three
dimensional space. In this space we fix an orthogonal reference frame, denoted as $(x, y, z)$. The vector $\lambda$ is assumed to be uniformly
distributed over the unit sphere, and will be uniquely specified by polar angles $\mu$ and $\tau$ which we unconventionally choose to take
values in the intervals $\mu \in [0,2\pi)$, $\tau \in [0,\pi)$. Such variables are related to the standard polar angles $\theta$ and $\phi$
according to \footnote{Note that, with this functional change, the surface element of the sphere turns out to be $d \Omega = \vert \sin{\mu}
\vert d \mu d \tau$}:

\begin{eqnarray}
  &\mu = \theta, \; \tau = \phi \quad &{\rm for} \; y \geqslant 0; \nonumber \\
  &\mu = 2 \pi - \theta, \; \tau = \phi - \pi \quad &{\rm for} \; y < 0.
\end{eqnarray}


The assignment of $\lambda = (\mu, \tau)$ uniquely determines the nonlocally possessed values (the certain measurement outcomes) of $A = {\bf a}
\cdot \sigma$ and $B = {\bf b} \cdot \sigma$ according to:
\begin{eqnarray}\label{defva}
a_{\psi}({\bf a},{\bf b},\mu,\tau) &=& {\rm sign} (\hat{\bf a} \cdot \lambda), \nonumber \\
b_{\psi}({\bf a},{\bf b},\mu,\tau) &=& - {\rm sign}  (\hat{\bf b} \cdot \lambda).
\end{eqnarray}

\noindent The vectors $\hat{\bf a}$ and $\hat{\bf b}$ lie in the plane identified by ${\bf a}$ and ${\bf b}$, and are obtained from these
vectors by rotating them in such a way that they are still symmetrically disposed with respect to the bisector of the angle $\omega$ (with
$0 \leqslant \omega \leqslant \pi)$ between ${\bf a}$ and ${\bf b}$, and $\hat{\bf a}$ and $\hat{\bf b}$ form an angle $\hat{\omega}$
satisfying, as in the case of Bell's model, $\hat{\omega} = \pi \sin^2{\frac{\omega}{2}}$. Notice that $\hat{\omega} \leqslant \omega$
when $\omega \leqslant \pi/2$, and $\hat{\omega} > \omega$ when $\omega > \pi/2$.

It turns out that our model is crypto-nonlocal, with the variables $\mu$ and $\tau$ playing the role of lower and upper level hidden
variables, respectively. In fact, from (\ref{defva}) it follows that every observable assumes the values $+1$ and $-1$ in opposite
hemispheres of the unit sphere of $\lambda$, which has uniform distribution. Integration over $\mu$ means integration over a maximal
circle, therefore $f_{\psi} ({\bf a}, \tau) = g_{\psi} ({\bf b}, \tau) = 0$, which, as already stressed, is a necessary
condition for crypto-nonlocality, in the case of the singlet.

To proceed, we evaluate now the averages on the variable $\mu$ of the correlation functions, i.e., expressions of the type:
\begin{equation}
E_{\psi,\tau}(A,B) = \frac{1}{4} \int_{0}^{2 \pi} a_{\psi}({\bf a},{\bf b},\mu,\tau) b_{\psi} ({\bf a},{\bf b},\mu,\tau) \vert \sin{\mu}
\vert d \mu.
\end{equation}
For simplicity, we will limit our consideration to $4$ directions in the ($x,z$)-plane such that:
\begin{eqnarray}
&&{\bf a} = (\sin{\alpha}, 0, \cos{\alpha}), \,\, {\bf a}^{\prime} = (- \sin{3 \alpha}, 0, \cos{3 \alpha}), \\
&&{\bf b} = (- \sin{\alpha}, 0, \cos{\alpha}), \,\, {\bf b}^{\prime} = (\sin{3 \alpha}, 0, \cos{3 \alpha}), \nonumber
\end{eqnarray}
determining the dichotomic observables $A$, $B$, $A'$ and $B'$, as usual.
To exhibit the emergence of quantum nonlocality in the standard scenario, and also of superquantum nonlocality, in our
context, it is sufficient to limit the attention to the interval $\alpha \in [0, \pi/4]$.
Let $\tilde{\alpha}$ be the solution of $4 \alpha + \widehat{2 \alpha} = \pi$, where $\widehat{2 \alpha} = \pi \sin^2{\alpha}$ (it turns out that
$\tilde{\alpha} \simeq 0.316$). For the aforementioned choice of physical settings,
we obtain $E_{\psi,\tau}(A,B) = 2 \vert \chi_1 \vert - 1$, $E_{\psi,\tau}(A^{\prime},B^{\prime}) = 2 \vert \chi_2 \vert - 1$,
\begin{equation}\label{aver}
  E_{\psi,\tau}(A,B^{\prime}) = E_{\psi,\tau}(A^{\prime},B) = \vert \chi_3 - \chi_4 \vert - 1
\end{equation}
when $0 \leqslant \alpha \leqslant \tilde{\alpha}$, and
\begin{equation}\label{aver2}
  E_{\psi,\tau}(A,B^{\prime}) = E_{\psi,\tau}(A^{\prime},B) = 1-  \vert \chi_3 + \chi_4 \vert \nonumber \\
\end{equation}
when $\tilde{\alpha} < \alpha \leqslant \pi/4$. In the above equations we have defined
\begin{equation}\label{expr1}
    \chi_j = \chi_j (\alpha,\tau) = \frac{\cos{\tau}}{\sqrt{\cos^2{\tau} + \cot^2{\frac{\gamma_j(\alpha)}{2}}}},
\end{equation}
with the functions $\gamma_j(\alpha)$ given by:
\begin{eqnarray}\label{gamma}
  &&\gamma_{1} (\alpha) = \pi \sin^2{\alpha}, \quad \gamma_{2} (\alpha) = \pi \sin^2{3 \alpha}, \\
  &&\gamma_{3} (\alpha) =  4 \alpha + \pi \sin^2{\alpha}, \quad \gamma_{4} (\alpha) =  4 \alpha - \pi \sin^2{\alpha}. \nonumber
\end{eqnarray}
We briefly describe how to derive the simplest joint correlation, $E_{\psi,\tau}(A,B)$. By using that
\begin{eqnarray}\label{details1}
    {\rm sign}(\hat{\bf a} \cdot \lambda) \, {\rm sign}(\hat{\bf b} \cdot \lambda) &=& {\rm sign}(\hat{\bf a} \cdot \lambda)
(\hat{\bf b} \cdot \lambda) \nonumber \\
&=& {\rm sign} (\chi^2_1 - \cos^2{\mu}),
\end{eqnarray}
we compute
\begin{equation}\label{details2}
     E_{\psi,\tau}(A,B) = \frac{1}{2} \int_0^{\pi} {\rm sign}(\chi^2_1 - \cos^2{\mu}) \sin{\mu} \, d \mu,
\end{equation}
from which the result follows. The other cases are rather cumbersome, but follow the same line of thought.

\noindent We combine now the joint correlations as usual, i.e., as in (\ref{f1}), with $E_{\psi,\tau}(X,Y)$
replacing $E(X,Y)$, getting
\begin{equation}\label{finalf}
     F_{\psi,\tau} = \left\{
                       \begin{array}{ll}
                         2 (\vert \chi_1 \vert - \vert \chi_2 \vert + \vert \chi_3 - \chi_4 \vert - 1), & \quad \alpha \in [0, \tilde{\alpha}], \\ \\
                         2 (\vert \chi_1 \vert - \vert \chi_2 \vert - \vert \chi_3 + \chi_4 \vert + 1), & \quad \alpha \in [\tilde{\alpha}, \frac{\pi}{4}].
                       \end{array}
                     \right.
\end{equation}
The function $F_{\psi,\tau}$ takes in general both positive and negative values lying in the interval $[-4, 4]$. However, it is easy to check
that, averaging this function over the remaining hidden variable $\tau$, we obtain the quantum mechanical expression
$F_{\psi}$. In fact,
\begin{equation}\label{check}
    \frac{1}{\pi} \int_0^{\pi} \chi_j (\alpha, \tau) \, d \tau = \frac{2}{\pi} \, \gamma_j (\alpha),
\end{equation}
producing $F_{\psi} = - {\bf a}\cdot{\bf b} - {\bf a}\cdot{\bf b}^{\prime} - {\bf a}^{\prime}\cdot{\bf b} + {\bf a}^{\prime}\cdot{\bf b}^{\prime}$,
which, as well known, saturates the quantum limit $2 \sqrt{2}$ at $\alpha = \pi/8$.

\begin{figure}[t]
\centering
\begin{center} 
 \includegraphics[width=8cm]{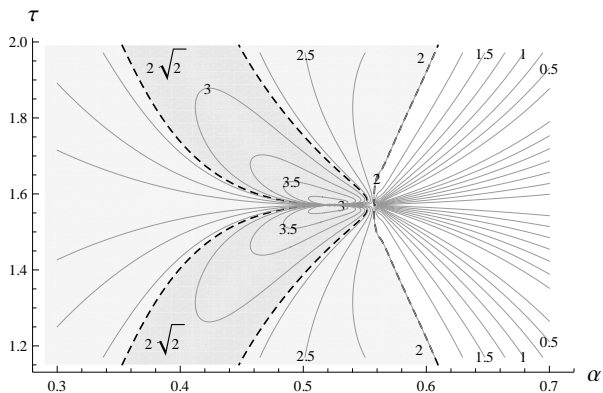} \\
 \caption{\footnotesize Contour plot of the function $\vert F_{\psi, \tau} \vert$, in the space of the parameters $(\alpha, \tau)$.
 This space is divided in three regions, corresponding to locality, $\vert F_{\psi, \tau} \vert \leqslant 2$, quantum nonlocality,
 $2 < \vert F_{\psi, \tau} \vert \leqslant 2 \sqrt{2}$, and superquantum nonlocality, $2 \sqrt{2} < \vert F_{\psi, \tau} \vert
 \leqslant 4$, identified by the dashed lines.}\label{fig1}
\end{center}
\end{figure}

Let us come back to study the function $\vert F_{\psi,\tau} \vert$, i.e. the modulus of the standard combination of values when the average is taken only on
the lower level variables. This function is continuous, exception made for a singular point at $(\alpha, \tau) = (\pi/6, \pi/2)$; in the
neighborhoods of this point the upper bound $4$ is approached,
\begin{equation}\label{supF}
    \sup \vert F_{\psi, \tau} \vert = 4.
\end{equation}
Summarizing, the correlations of the simple model we have presented, when only the averages over the lower level variables are
taken into account, approach, for  values of the variable $\tau$ in an appropriate interval, the maximal possible violation of
locality in a two-qubits system. Therefore, our model, which is physically  meaningful, as shown by its similarity with standard hidden variables models, is not formulated purely in formal mathematical terms and still it is able to simulate the Popescu-Rohrlich box from the point of view of its violation of locality.
We can partition the space of parameters $(\alpha, \tau)$ in three regions, corresponding to
locality, quantum nonlocality, and superquantum nonlocality, see Fig. \ref{fig1}.

{\it Conclusions ---} We have developed a simple nonlocal model for the singlet state of the two-qubits system, which is
predictively equivalent to quantum mechanics when one integrates over all the hidden variables, while exhibiting
superquantum nonlocality at the level of the lower hidden variables. This model is strongly based on the
famous model of J. Bell, and differs from it only by the way in which nonlocality enters the description, and by the fact that
a partial integration over the hidden variables already erases nonlocality, when dealing with local observables.

The main finding of this paper is that nonlocal hidden variables theories of this type, at the intermediate level
can exhibit the largest possible nonlocality, in a bipartite system of two qubits. In our
model, the upper bound $4$ is  approached only asymptotically, but we believe that models in which this value
is effectively attained are possible.

As a consequence of our analysis, the PR device, and more generally any device exhibiting superquantum correlations, is not only
a mathematical tool. There is indeed a strong connection between hidden variables theories and superquantum correlations. In their
inspiring work, Rohrlich and Popescu were looking for a physical principle,  relativistic causality, which could justify the
precise way nonlocality enters the quantum mechanical description. They concluded that this was not the case. Our analysis suggests
that this idea can be revitalized in the domain of crypto-nonlocal models, which
deserve a particular attention in the identification of the physical principles underlying nonlocality.


\end{document}